\RequirePackage{latexsym}%
\RequirePackage{amssymb}%

\input{aipcheck}

\documentclass[
    ,final            
  ]
  {aipproc}

\layoutstyle{8x11double}

\newcommand{\feoh}{[{\rm Fe} / {\rm H}]}
\newcommand{\msun}{\, M_\odot}
\newcommand{\mmd}{M_{\rm md}}

\newcommand{\nucm}[2]{${}^{#1}{\rm #2}$}
\newcommand{\abra}[2]{[{\rm #1}/ {\rm #2}]}

\newcommand{\apj}{{\it{ApJ}}}

\newcommand{\aj}{{\it{AJ}}}
\newcommand{\aap}{{\it{A\&A}}}
\newcommand{\mnras}{{\it{MNRAS}}}

\begin{document}

\title{The IMF of Extremely Metal-Poor Stars \\ and the Probe into the Star-Formation Process \\ of the Milky Way}

\classification{<Replace this text with PACS numbers; choose from this list:
                \texttt{http://www.aip..org/pacs/index.html}>}
\keywords      {metal-poor star, carbon star, star formation, star abundance, binary, galaxy formation, galaxy merger, our Galaxy}

\author{Yutaka Komiya}{
  address={Department of Astronomy, Faculty of Science, Tohoku University, Sendai, Miyagi Prefecture 980-8578, Japan}
}

\author{Takuma Suda}{
  address={Department of Physics, Faculty of Science, Hokkaido University, Sapporo, Hokkaido 060-0810, Japan}
}

\author{Asao Habe}{
  address={Department of Physics, Faculty of Science, Hokkaido University, Sapporo, Hokkaido 060-0810, Japan}
}

\author{Masayuki Y. Fujimoto}{
  address={Department of Physics, Faculty of Science, Hokkaido University, Sapporo, Hokkaido 060-0810, Japan}
}

\begin{abstract}
 We discuss the star formation history of the Galaxy, based on the observations of extremely metal-poor stars (EMP) in the Galactic halo, to gain an insight into the evolution and structure formation in the early universe.  
   The initial mass function (IMF) of EMP stars is derived from the observed fraction of carbon-enhanced EXP (CEMP) stars among the EMP survivors, which are thought to originate from the evolution in the close binary systems with mass transfer. 
   Relying upon the theory of the evolution of EMP stars and of their binary evolution, we find that stars of metallicity $\feoh \lesssim -2.5$ were formed at typical mass of $\sim 10 \msun$. 
   The top heavy IMF thus obtained is applied to study the early chemical evolution of the Galaxy.  
   We construct the merging history of our Galaxy semi-analytically and derive the metallicity distribution function (MDF) of low-mass EMP stars that survive to date with taking into account the contribution of binary systems.  
   It is shown that the resultant MDF can well reproduce the observed distribution of EMP survivors, and, in particular, that they almost all stem from a less-mass companion in binary systems.  
   We also investigate how first stars affect the MDF of EMP stars. 
\end{abstract}

\maketitle


\section{Introduction}
Extremely metal-poor (EMP) stars in our Galaxy are the relics of the early universe and can be a useful probe into the structure formation through merging history and into the first stars. 
   Thanks to the large-scale survey of metal-poor stars in the Galactic halo such as HK survey\citep{Beers92} and Hamburg/ESO (HES) survey\citep{Christlieb01}, thousands of stars are found with $\feoh<-2$. 
   Such a large number of EMP stars enable us to discuss trends in the chemical composition of these stars with use of the follow-up observations of high dispersion spectroscopy. 
   Among them, one of the known features of EMP stars is that a large fraction ($\gtrsim 20$ \%) of them show the enhancement of carbon, by far larger than the fraction (a few \%) that the carbon stars occupy among Population I and II stars. 
   Carbon-enhancement can be associated with the mass transfer from the AGB companion in close binary systems as in the case for metal-rich carbon stars such as CH- and Ba-stars, attendant also with the enhancement of s-process elements. 
   It is natural that we connect the origin of carbon-enhanced EMP (CEMP) stars with the survivors of low-mass members in the binary system whose companions were once on the AGB and experienced dredge-up of nuclear products. 
   Indeed, most of CEMP stars ($\sim 70$ \%) show enhancement of s-process elements like barium and lead, although others show normal abundance of barium. 
   The former is called CEMP-s, while the latter CEMP-no or CEMP-nos.

It is known that the stars of extremely low- and zero-metallicity show peculiar evolution, inherent in low- and intermediate-mass models.  
  Because of initial smaller abundances of CNO elements, the entropy in the hydrogen burning shell is smaller, and hence, the helium flash-convection extends through the tail of hydrogen-rich layer, differently from the stars of younger generations \citep{Fujimoto90}.  
   Then, engulfed hydrogen is mixed inward and burns in the middle of convection to ignite a hydrogen shell-flash, and during the decay phase, the surface convection deepens to carry out CN elements and s-process elements to enrich the surface. 
   It is shown that this so-called ``Helium-Flash Driven Deep Mixing" (He-FDDM) occurs at core helium flash on the red giant branch or at helium shell flash on the TP-AGB phase for $\feoh \lesssim -2.5$ and $M \lesssim 3.5 \msun$ \citep{Hollowell90,Fujimoto00,Iwamoto04}. 
   The low-mass members in the binary systems with the primary stars in these metallicity and mass range will show enhancement of these elements and becomes CEMP-s stars after they undergo the mass transfer from the AGB companions. 
   \citet{Suda04} have applied the models with the He-FDDM to interpret the origin of star HE0107-5240, the most iron-deficient star at that time. 

For stars in the mass range of $3.5 \lesssim M / \msun \lesssim 6$, on the other hand, hydrogen mixing, and hence, the He-FDDM, do not occur but the third dredge-up operates to enrich the surface with carbon as the stars of younger generations. 
   In this case, however, the enrichment of s-process elements, as suggested from the observations is not necessarily accompanied \citep{Komiya07}. 
   Accordingly, the secondary stars with primary star in this range may show up as CEMP-nos stars after the mass transfer;  
   the upper limit of the primary mass range is imposed by the hot bottom burning in the envelope, which converts the dredged carbon into nitrogen.  

In this work, we first explore the initial mass function (IMF) in the early universe based on the binary scenario for the subclasses of CEMP stars, i.e., CEMP-s and CEMP-nos stars. 
   Then we Apply the IMF thus obtained to figure out the metallicity distribution function (MDF) and compare it the observed distribution of EMP stars in the Galactic halo. 
   In deriving the MDF, we study the chemical evolution of the Galaxy with take into account the merging history of minihalos.  
   In this work, we define EMP and CEMP stars as those of $\feoh \le -2.5$ and the EMP stars with $\abra{C}{Fe} \ge 0.5$, respectively; 
   CEMP-s and CEMP-nos are the subclasses of CEMP, divided according to whether a star has $[s /{\rm Fe}] \ge 0.5$ or not, where $[s /{\rm Fe}] $ denotes the enhancement of s-process elements such as Ba and Pb. 
   We exclude stars with the enhancement of r-process elements from the group of CEMP-s.  


\section{IMF of EMP stars} 

The IMF can be determined from the theory of stellar evolution and of binary mass transfer once the fraction of the number of CEMP stars among EMP stars, i.e., $\sim 10-20$ \%, and the ratio of the fraction of CEMP-s stars to that of CEMP-nos stars, i.e., $\sim 0.33$ \citep{Aoki02} are given. 
   As stated above, the existence of CEMP stars are associated with the He-FDDM at RGB or AGB as well as the third dredge-up (TDU). 
   For EMP stars, the occurrence of TDU depends on the initial mass and metallicity of a star as discussed in this section. 
   On the other hand, the production site of s-process elements for EMP stars are assumed to take place in the helium flash convective zone of primary stars that experience the He-FDDM during thermally pulsating AGB phase. 
   We assume the standard \nucm{13}{C} pocket not to work under the EMP circumstance, while  this assumption may demand in more detail discussion. 
   Accordingly, CEMP-s stars are considered to be the outcome of He-FDDM and/or TDU in low- and intermediate-mass stars, while CEMP-nos to experience only the TDU in the models of more massive AGB stars. 
   The final fates of EMP stars in the various mass range are discussed in this conference proceedings \citep{Suda07}. 
   One should note that massive CEMP-nos stars may also undergo the hot bottom burning in the convective envelope.  
   Such stars are recognized as nitrogen-rich EMP stars \citep{Spite05,Sivarani06}. 
   Since these stars are comparable, or even exceed, the CEMP-nos stars in number, the above ratio of CEMP-s to CEMP-nos is an upper limit.

The observed fraction of CEMP, and relative frequencies of CEMP-s and CEMP-nos stars, can be used as the constraint on the IMF of EMP stars by assuming that all of these class of stars were born as the members of binary system whose primary stars have mass in the range of $\sim 1-7 \msun$. 
   We assume the IMF of Millar-Scalo type lognormal function in the form;
\begin{equation}
\xi (m)\propto \frac{1}{m} \exp \left( -\frac{(\log (m/ \mmd))^2}{2\times \Delta _M^2} \right). 
\label{eqIMF}
\end{equation} 
   where $\mmd$ and $\Delta _M$ is the medium mass and dispersion of the mass function, respectively. 
   From the theory of stellar evolution, we find the upper and lower mass limits, $M_{up}$ and $M_{low}$, to the occurrence of He-FDDM and TDU, and also the upper and lower limits, $d_{up}$ and $d_{low}$, of binary separations that the low-mass member can accrete the gas from the AGB companions. 
    Then, with relevant values of these upper and lower limits, the fraction of the stars of these groups are given as follows:
\begin{equation}
\psi \propto \int_{M_{low}}^{M_{up}} \xi (m) \frac{n( M_{2} / m )}{m} dm \\
\times \int_{d_{low}}^{d_{up}} f(P) \frac{dP}{da} da
\end{equation} 
   where $f(P)$ and $n(q)$ is the period distribution and mass-ratio function of the binary, respectively. 
   Here, we choose $n(q) = 1 (q = m_{2} / m_{1} )$ for simplicity and $f(P)$ is taken from \citet{DM91} for stars in the Galactic bulge. 
   We may set the range of the first integral between $M_{low} = 0.8 \msun$ and $M_{up} = 3.5 \msun$ in which the He-FDDM occurs for the fraction of CEMP-s stars, for example. 
   The second integral gives the range of binary separation that allows the secondary stars to gain the amount of carob-rich matter from AGB companion large enough to be observed as CEMP stars. 
   The lower bound is determined by the radius of the star when it experience He-FDDM and/or TDU on the AGB, while the upper bound is determined by the binary separation that allow the secondary to attain the surface carbon enrichment $\abra{C}{Fe} \ge 0.5$ through the accretion from the wind of AGB stars and despite of the dilution by the surface convection on the secondary. 

\begin{figure}[btp]
  \includegraphics[width=\columnwidth]{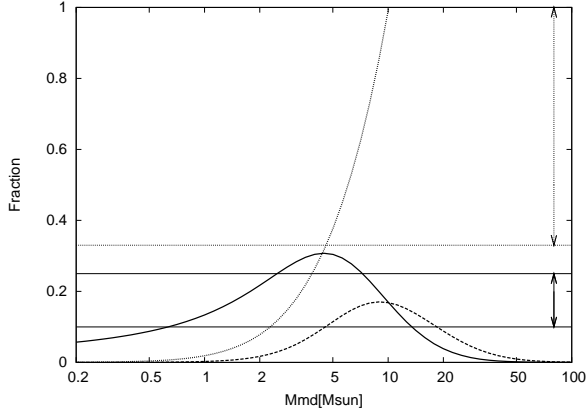}
  \caption{Fractions of CEMP-s and CEMP-nos stars among EMP stars as a function of the medium mass $\mmd$, a parameter of the IMF in eq.~(\ref{eqIMF}), with the other parameter fixed at $\Delta _M = 0.33$. 
   Solid and broken lines plot the proportions of CEMP-s and CEMP-nos stars to EMP stars, predicted from the binary scenario, respectively. 
   Dotted line denotes the number ratio of CEMP-nos to CEMP-s stars.   
   The observed proportion of CEMP-s stars ($\simeq 10 \sim 25 \%$) and the ratio between CENP-nos and CEMP-s stars ($\simeq 1/3 \sim 1$) are bounded by solid and dashed horizontal lines and two arrows indicate the ranges compatible with the observations.  
}
  \label{Mmd}
\end{figure}

The predicted ratios of the fraction of CEMP-s, and CEMP-nos relative to the fraction of binaries are shown as a function of $\mmd$ of the IMF in Figure~\ref{Mmd}. 
   In order to explain the consistent fraction of these group of EMP stars, $\mmd \simeq 10 \msun$ is required, although $4-6 \msun$ is satisfactory if we consider the fraction of CEMP-s stars alone. 
   In conclusion, the IMF of the EMP stars are implied to be quite different from the present-day IMF that peaks at a low mass. 
   This result also suggests that the most of the observed CEMP stars are the relics of the secondary members in the binary systems because the contribution of low-mass stars, born as single, is very small in such top-heavy IMF.

\section{Early Metal Enrichment of our Galaxy}

In this section, we estimate the MDF of the Galaxy and compare with the observed MDF among EMP stars by applying the IMF of EMP stars derived above. 
   Since the metal enrichment of the Galaxy is caused by massive stars that explode as Type II supernovae, we have to estimate the ejected mass from supernovae in the early phase of the Galaxy. 
   Let us suppose that the Galaxy of mass $M_{h}$ has the $N(Z)$ stars, where Z is the average metallicity of the Galaxy at a given time. 
   If $\Delta N(Z)$ of new stars are born from the gas clouds and the average metallicity increases by $\Delta Z$, the increment in the mass of metals by SNe is given by,
\begin{equation}
M_h\Delta Z = \int_{M_{SN}}^{\infty}  Y(m)\xi_t(m)dm \Delta N(Z). 
\end{equation}
where $M_{SN}$ is the lower mass limit of stars that explode as Type II SNe, $\xi_{t}$ is the time dependent IMF, normalized as $\int_{0}^{\infty}\xi_t(m)dm = 1$, and $Y(m)$ is the ejected mass of iron as a supernova yield from stars of initial mass $m$ and set at $0.07 \msun$ as a average in this study. 
   For simplicity, we assume that the supernova ejecta are instantaneously distributed in the whole Galaxy and that the chemical composition of the next generation stars are homogeneous.  
   With use of the above equation, the number $n(Z)$ of EMP survivors at metallicity $Z$ is
\begin{equation}
n(Z) = \frac{dN(Z)}{dZ} \times \int_{0.08}^{0.8} \xi_t (m) dm
= M_h \frac{\int^{0.8}_{0.08} \xi_t (m)dm}{\int_{M_{SN}}^\infty  Y(m) \xi_t(m) dm}. 
\end{equation} 
   It is worth note that $n(Z)$ is independent on $Z$ because $\xi_{t}(m)$ is set to be independent of $Z$ in this study. 
   When we change the variable for $n$ from $Z$ to \feoh by simply setting $Z = Z_{\odot} 10^{\feoh}$, 
\begin{equation}
n(\feoh ) = n(Z) \frac{dZ}{d\feoh} = n(Z) \ln(10) Z_\odot 10^{\feoh} .
\end{equation} 
   This gives the theoretical MDF for EMP stars in the whole Galactic halo.

Using the above MDF, we can compare it with the observed MDF in the Galactic halo, covered by HES survey. 
   In order to obtain the partial MDF in the limited sample volume by the survey, we assumed that (1) the HES sample includes all EMP giants in the Galactic halo within the field of view because the limiting magnitude of the survey is large enough to reach farthest giants, (2) the HES sample includes the same number of EMP dwarfs as derived from the total number of dwarfs estimated from the number giants by considering the relative life times and with the stellar luminosity and the flux limit of survey taken into account, and (3) the spatial distribution of halo EMP stars is isotropic.  

\begin{figure}[htbp]
  \includegraphics[width=\columnwidth]{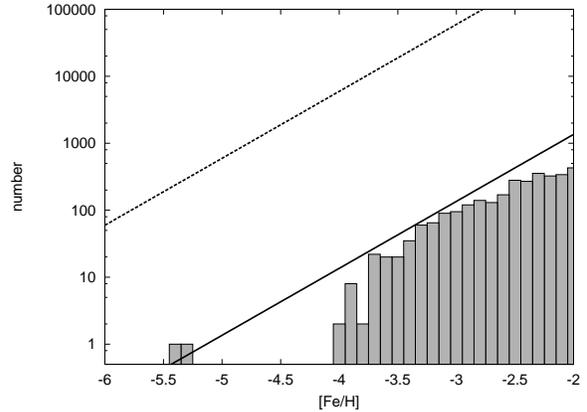}
  \caption{Metallicity distribution function of EMP stars. 
   Shaded histogram shows distribution of observed EMP stars by HES survey \citep{Beers05}. 
   Two straight lines denote the predictions from simple one zone model;  
   solid line for the IMF with $M_{md} = 10\msun$ and thin broken line for the same IMF as the Galactic spheroid. 
   MDF with the top heavy IMF give a reasonable fit to the observations except for the metallicity range below $\feoh \lesssim -4$. 
}
  \label{MDF1}
\end{figure}

Figure~\ref{MDF1} shows the comparison with the observed MDF.  
   It reveals that the MDF derived from the top heavy IMF of $M_{md} \simeq 10 \msun$ is compatible with the observed one, while the IMF peaked at low-mass significantly overestimates the number of observed EMP survivors.  
   This is simply because the low-mass peaked IMF produces long-lived, low-mass stars much more than the top heavy IMF while producing necessary amount of metals. 
   For high-mass peaked IMF, the number of low-mass survivors are smaller by a factor of 1,000 than for the low-mass peaked IMF. 
   There are two factors. 
   One is the small number of low-mass stars born, which decreases the numerator in eq.~(4). 
   The other is the large metal-enrichment from a single generation of stars, which increases the denominator in the equation.  
   The fact that the observed MDF in the Galactic halo is well reproduced for $\feoh \gtrsim -4$ implies that the early phase of the chemical evolution of our Galaxy gives an additional support to the top heavy IMF, derived above from the observed ratio of CEMP-s and CEMP-nos. 
   It is to be noted, however, that the sharp cutoff at $\feoh \sim -4$ is hardly to be explained within this simple one zone model of chemical evolution and this problem is discussed in the next section.

\section{Merging History and Formation of EMP Stars}

The observed cutoff of MDF at $\feoh \lesssim -4$ and two known stars at $\feoh < -5$ suggest that they belong to the separate population whose origin is different. 
   In particular, the former can be taken as evidence of the hierarchical structure formation process in the early universe. 
   In the current understandings of the hierarchical clustering in $\Lambda$ CDM universe, the typical mass of the primordial gas cloud is $\sim 10^5 \msun$, embedded in the dark matter mini-halo of mass of $\sim 10^6 \msun$ \citep{Tegmark97,Yoshida03}. 
   The currently observed MDF should reflect the merging history of the gas clouds starting from the primordial cloud totally devoid of metals. 
   During the formation process of the Galaxy, the gas clouds are polluted with metals by the stellar yields from massive stars.  

With the current knowledge of cold-dark matter scenario, we can follow the formation process of the Galaxy using the merging tree of mini-halos starting with the primordial gas clouds. 
   We may construct the merging tree program by using the theory of extended Press-Schechter approach \citep{LC93, Bond91}. 
   In developing a merger tree, we adopted a method of \citet{SK99}.  
   We set the halo mass of $M_h =10^{12} \msun$ for the final stage and of $M_l$ for the minimum mass of star forming halo by assuming that the virial temperature attains $T_{vir}(M_l) = 10^3K$ \citep{Tegmark97}. 
   We also include the effect of chemical evolution with the metal-pollution of gas clouds under the assumption of one-zone approximation.  
   Thus the gas cloud is polluted by a single supernova to be enriched with metals as much as $\feoh = -3.5$. 
   In the subsequent generations of clouds, stars are formed with polluted gas. 
   If the polluted mini-haloes are merged with the primordial clouds, the metallicity in the massive mini-halo can be less than -3.5 in \feoh. 
   In the primordial clouds of small masses, stars are also affected by the accretion from the interstellar gas polluted with metals due to their low virial velocity. 
   We treat the accretion process according to the Bondi's formalism with a given virial velocity and the stellar mass of $0.8 \msun$. 
   Depending on the merging history of individual clouds, the stars born in the primordial mini-haloes can be polluted up to $\feoh \sim -5$ through the accretion of interstellar gas. 
   More details on the program are described at Komiya et al. (2007, in preparation).

\begin{figure}
  \includegraphics[width=\columnwidth]{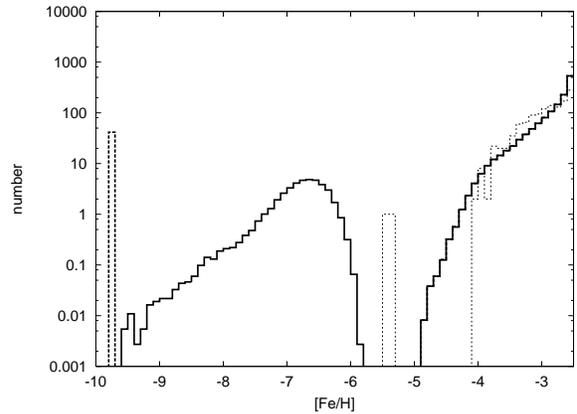}
  \caption{Metallicity distribution function of EMP stars. 
\textit{Thick solid line}: Predicted MDF resulting from merger tree with the surface pollution taken into account. 
\textit{Thick dashed line}: Predicted distribution of pristine metallicity. 
\textit{Thin dotted line}: Distribution of observed EMP stars by HES survey \citep{Beers05}. 
   Predicted MDF is consistent with observation of EMP stars, but inconsistent for HMP stars of $\feoh <-4$. 
}
  \label{MDF2}
\end{figure}

\begin{figure}
  \includegraphics[width=\columnwidth]{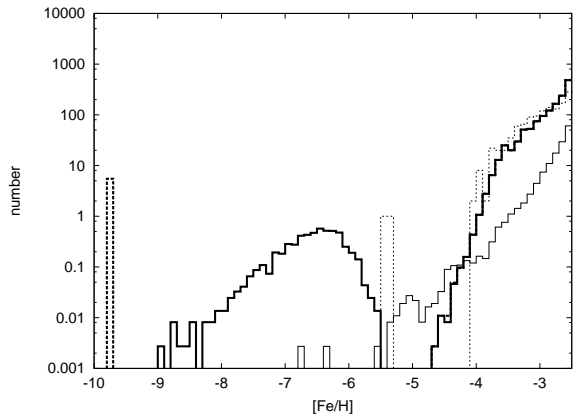}
  \caption{
Same as fig. \ref{MDF2}, except for stars formed in zero metal gas. 
\textit{Thick lines}: Case of $M_{\rm md} = 100\msun$ in IMF for zero metal gas, and later $M_{\rm md} = 10 \msun$.  Solid and dashed line shows MDF after and before surface pollution, respectively. 
  The predicted number of Pop.~III stars is comparable to observed number of HMP stars. 
  \textit{Thin solid line}: All Pop.~III stars assumed to explode as PISNe and all gas to escape from mini-halo. 
  The predicted number of EMP stars is much fewer than the observed number because PISN eject larger amount of metals. 
   Thin dotted line shows the observed MDF. 
}
  \label{100}
\end{figure}

The results of the computations are illustrated in Figures~\ref{MDF2} and \ref{100}. 
   Fig.~\ref{MDF2} show the results obtained by assuming the IMF of $M_{\rm md} = 10 \msun$.  
   Clearly, we can see the two distinct groups of stars: 
   the lower metallicity group comes from the first generation stars, while the more metal-rich group comes from the second and subsequent generations. 
  The predicted MDF well reproduces the sharp drop at $\feoh \lesssim -4$, although the cutoff may is not sharp enough compared with the observation. 
  Serious discrepancy is found for the population of primordial stars with metal pollution by accretion. 
  Since the merging trees starts from a large number of primordial clouds, low-mass primordial stars are significantly overproduced and survive to date in the Galactic halo.

Fig.~\ref{100} shows the results for different IMF of $M_{md} = 100 \msun$.
This is the consideration of the suggestion from the star formation theory that primordial clouds prefer to be born as very massive stars as much as $\sim 100 \msun$.
This result denoted by thick solid line fits better to the observed MDF, althogh the width of the distribution function for Pop.III is still inconsistent.
For very top heavy IMF, the contribution of pair-instability supernova (PISN) in the mass range of $150-280 \msun$ becomes large.
The thin solid line in Fig.~\ref{100} is the result of the simulation if all Pop.III are explode as PISNe and eject $10 \msun$ of iron per event.
Since the velocity of PISN is very large enough to exceed the escape velocity of mother clouds, the ejected mattar are dispersed into the halo and enrich the metallicity of other clouds.
As a result, the Pop.III survivors become metal-rich and merged into the EMP population.
The slope of the MDF also becomes less steep due to the rapid increase of metallicity in the final halo.
In conclusion, the massive IMF is prefereble in reproducing the observed MDF, but the contribution of PISN should be negligible or PISN do not work in the early universe.

\section{Summary}

We discuss about the star formation history of our Galaxy using the observation of EMP stars with the current knowledge of the stellar evolution, and within the framework of hierarchical clustering in $\Lambda$ CDM universe.  
From the comparison of observed CEMP, CEMP-s, and CEMP-nos stars with the binary scenario, the currently observed EMP survivors were born in the binary system with the IMF of the typical mass of $10\msun$. 
  We have derived the metallicity distribution function based on the simple one zone model of chemical evolution to find that the derived top heavy initial mass function is consistent with the observed metallicity distribution function. 
   We also construct the merging tree of mini-halos starting with the primordial gas clouds of $\sim 10^5 \msun$ until the formation of the present Galactic halo of $10^{12} \msun$. 
   The predicted metallicity distribution function with merging history can reproduce a sharp cutoff around $\feoh \sim -4$, but is subject to significant overproduction of Pop.~III objects, not supported by the observations. 
   The inclusion of the contribution of pair-instability supernovae is not desirable in reproducing the metallicity distribution function of EMP halo stars due to their large amount of iron ejecta.

Although this work is an important step, more sophistication of the models is desirable for better understanding of the star formation history of the Galaxy.
   The next generation of the project of the search for EMP stars in the Galaxy (SDSS/SEGUE, LAMOST) will provide more insight into the understandings of the structure formation of the Galaxy, leading us to the field of ``near-field cosmology''.



\bibliographystyle{aipproc}   


\IfFileExists{\jobname.bbl}{}
 {\typeout{}
  \typeout{******************************************}
  \typeout{** Please run "bibtex \jobname" to optain}
  \typeout{** the bibliography and then re-run LaTeX}
  \typeout{** twice to fix the references!}
  \typeout{******************************************}
  \typeout{}
 }


\end{document}